# A Survey of Reverse Engineering and Program Comprehension

Michael L. Nelson
<*m.l.nelson@larc.nasa.gov*>

## Abstract

*Reverse engineering has been a standard practice in the hardware community for some time. It has only been within the last ten years that reverse engineering, or "program comprehension," has grown into the current sub-discipline of software engineering. Traditional software engineering is primarily focused on the development and design of new software. However, most programmers work on software that other people have designed and developed. Up to 50% of a software maintainers time can be spent determining the intent of source code. The growing demand to reevaluate and reimplement legacy software systems, brought on by the proliferation of client-server and World Wide Web technologies, has underscored the need for reverse engineering tools and techniques. This paper introduces the terminology of reverse engineering and gives some of the obstacles that make reverse engineering difficult. Although reverse engineering remains heavily dependent on the human component, a number of automated tools are presented that aid the reverse engineer.*

## Introduction of and Motivation for Reverse Engineering

"Reverse engineering" has its origins in the analysis of hardware for commercial or military advantage [4]. The purpose is to deduce design decisions from end products with little or no additional knowledge about the procedures involved in the original production. The same techniques are currently being researched for application to legacy software systems, not for industrial or defense ends, but rather to recover incorrect, incomplete, or otherwise unavailable documentation.

Software reverse engineering, or program comprehension or understanding, is a research area devoted to developing tools and methodologies to aid in the understanding and management of the increasing number of legacy systems. Traditional software engineering research and development focuses on increasing the productivity and quality of systems under development or being planned [24].

However, maintenance of existing systems is estimated to consume between 50% and 80% of the resources in the total software budget [2, 15]. Within the maintenance function, reverse engineering activities ("comprehension") require 47% and 62% of the total time for enhancement and correction tasks, respectively [7].

Without diminishing the importance of software engineering activities focusing on initial design and development, empirical evidence suggests that significant resources are devoted to reversing the effects of poorly designed or neglected software systems. In a perfect world, all software systems, past and present, would be developed and maintained with the benefit of well structured software engineering guidelines. In the real world, many systems are not or have had their structured design negated, and there must be tools and methodologies to handle these cases.

## Reverse Engineering Defined

To better define the area of reverse engineering, it is first necessary to explain in the larger context of the software system lifecycle. The following terms and definitions are adapted from the canonical taxonomy given in [4]:

> Forward Engineering: This term, the obvious opposite of reverse engineering, is offered to distinguish the traditional software engineering process from reverse engineering.
>
> Reverse Engineering: The process of identifying software components, their inter-relationships, and representing these entities at a higher level of abstraction. Reverse engineering by itself involves only analysis, not change. Program comprehension and program understanding are terms often used interchangeably with reverse engineering Four specializations of reverse engineering are offered, in increasing level of impact:
>
>> *Redocumentation:* Perhaps the weakest form of reverse engineering, this involves merely the creation (if none existed) or revision of system documentation at the same level of abstraction.
>>
>> *Design Rediscovery:* Redocuments, but uses domain knowledge and other external information where possible to create a model of the system at a higher level of abstraction.
>>
>> *Restructuring:* Lateral transformation of the system within the same level of abstraction. Also maintains same level of functionality and semantics.
>>
>> *Reengineering:* The most radical and far reaching extension. Generally involves a combination of reverse engineering for comprehension, and a reapplication of forward engineering to reexamine which functionalities need to be retained, deleted or added.
>>
>> Substituting "business process" or "organization" for "software system" is the subject area of the popular Business Process Reengineering [10, 11]. The techniques are the same, just applied to a large scope.

Figure 1 illustrates the relationship between forward and the degrees of reverse engineering.

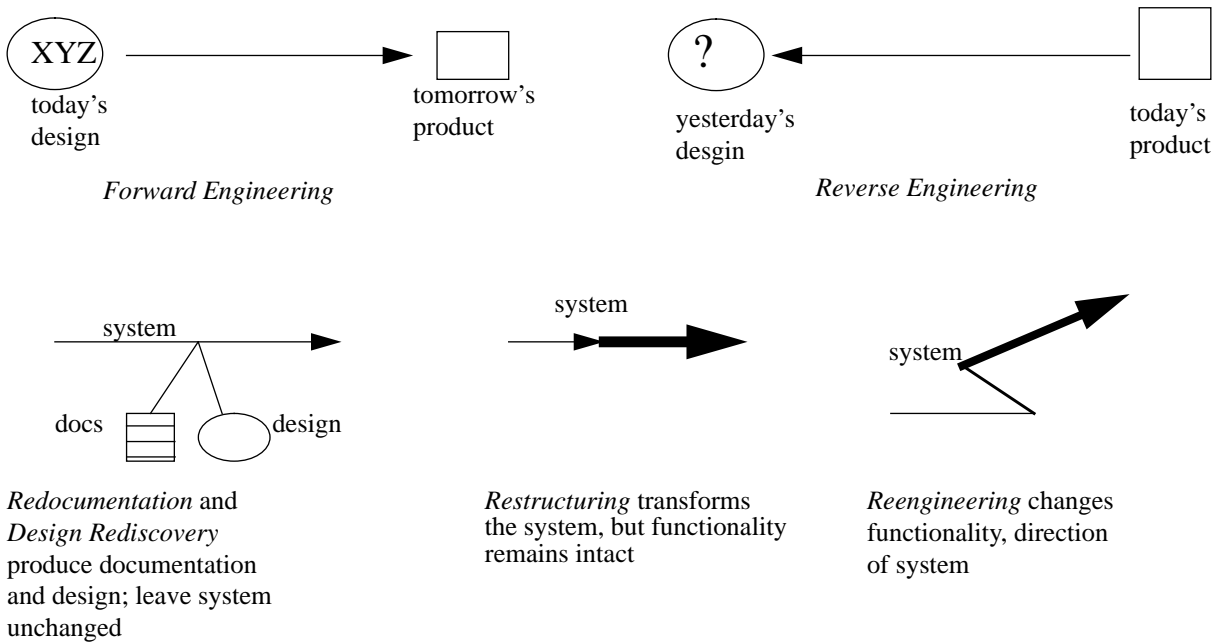

Figure 1: Forward Engineering, Reverse Engineering, and Derivatives

## Difficulties for the Reverse Engineer

Reverse engineering is a challenging task because it involves mapping between different worlds in five distinct areas [21, 24]:

1. *Application Domain <--> Programming Language*
A programming language is just a model environment to solve some real problem. While tools exist to assist in understanding what the code is doing from a code perspective, there is little to assist the reverse engineer in determining what is occurring with the code from a domain perspective.

2. *Machines and Programs <--> Abstract, High-Level Design*
Simple, abstract concepts ("sort the list of customers by last name") quickly become lost in the minutia detail of programming. Computer science education is largely about mapping from the abstract to the detailed implementation, but there is little to assist in the reverse mapping.

3. *Original Coherent, Structured System <--> Actual System, With Structure Decaying*
Even when good documentation is available for a system, maintenance over time causes the structure to drift from the original specification [1]. The reverse engineer must be able to reconcile and synchronize the documented design and the current implemented design.

> 4. *Hierarchical Programs <--> Cognitive Association*
> Computer programs and formal, hierarchical expressions. Humans think in associative "chunks" of data. A reverse engineer must be able to "build up correct high level chunks from the low level details evident in the program" [24].
>
> 5. *Bottom-Up Code Analysis <--> Top-Down Application Analysis*
> Code analysis is by its nature a bottom-up exercise. It requires, simultaneously, higher level meaning to be extracted from code fragments, and higher level concepts to be mapped to lower level implementations. To make this task even more difficult, the engineer must be able to handle obfuscations such as *interleaving* [23]. Interleaving is the intentional (for optimization) or accidental (poor design or sloppy maintenance) co-location of logically separate tasks within the same spatial sequence.

Currently, reverse engineering is currently heavily dependent on human interaction and steering [enc]. While there are tools to assist the reverse engineer in program comprehension, it is not a fully automated process. The human element present in program comprehension is the subject of another field, *software psychology*, pioneered by Shneiderman [25]. Software psychology measures human performance while interacting with computer and information systems. For surveys on software psychology, the reader is referred to [16, 19].

## Approaches to Automating Reverse Engineering

A variety of approaches for automated assistance are available for the reverse engineer in program comprehension. A full list of reverse engineering approaches is available in [24]. Some of the more prominent approaches include:

> 1. *Textual, lexical and syntactic analysis* - these approaches focus on the source code itself and its representations. These include the use of UNIX's lex, lexical metrics (counting assignments, identifiers, etc.) outlined in [9], and even automated parsing of the code searching for cliches [26]. Cliches are standard approaches to problem solving that can extracted from the source code to give hints about design decisions. The unit of examination is the program source itself.
>
> 2. *Graphing methods* - there are a variety of graphing approaches for program understanding. These include, in increasing order of complexity and richness: graphing the control flow of the program[12], the data flow of the program [12], and program dependence graphs [6]. The unit of examination is a graphical representation of the program source.
>
> 3. *Execution and testing* - there are a variety of methods for profiling, testing, and observing program behavior, including actual execution and inspection walkthroughs. Dynamic testing and debugging is well known and there are several tools available for this function. For large systems, a technique called "partial evaluation" is available to identify and test isolate components of a system [17]. "Abstract interpretation" is a method for using denotational semantics to perform static testing through simulating the behavior of the actual system [5]. The unit of examination is a full, partial, or simulated execution of the program.

## Synchronized Refinement

A representative hybrid method of program understanding is known as *Synchronized Refinement*. "Synchronized Refinement consists of two parallel activities: the synthesis of functional and non-functional behavioral descriptions and the code-level analysis of the program text" [18]. This parallel relationship is depicted in figure 2.

Synchronized Refinement is an iterative process, using automated tools for code-analysis and human "steering" for course correction. Initial supposed design decisions, acquired from documentation, domain knowledge, or some additional non-source code information source, seed the process. Then detailed source code analysis is performed to verify the proposed design decision against observed implementation. Differences are noted and the design decisions revised. With a more accurate model of the design decisions, the process is repeated again, continuously refining and adjusting the direction of the analysis.

The developers of Synchronized Refinement report their results on a number reverse engineering projects, including military, telephony, and real-time systems, as well as the tools developed to assist in Synchronized Refinement in [18, 14, 20].

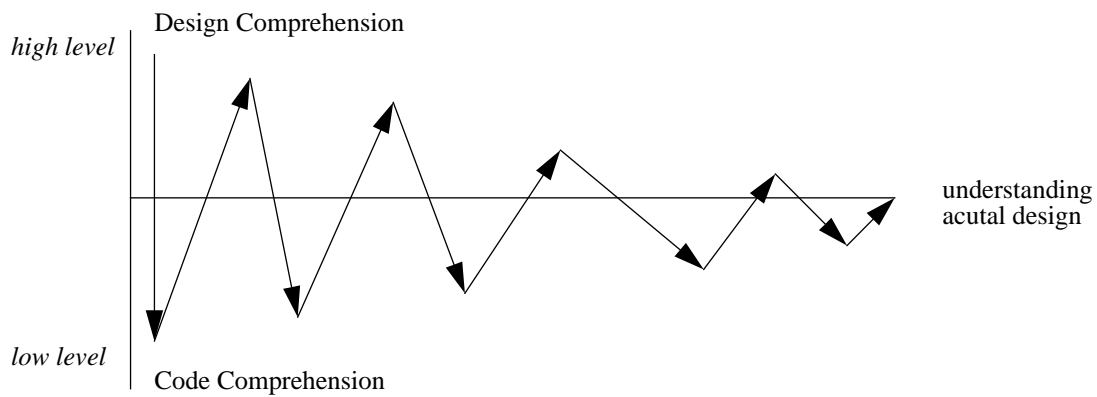

Figures 2: Synchronized Refinement iterating to actual design

## Application to Legacy Systems

Applying reverse engineering to legacy systems is a growth business. According to [27], there are hundreds of billions of lines of source code in the world, and 70% of it is COBOL. Within the realm of scientific computing, the code is almost entirely FORTRAN [13]. It takes little imagination to speculate on the condition and style of the code, the presence and quality of documentation, and lack of corporate memory of the understanding how these codes actually work.

In some cases, organizations consider software systems "capital assets" because they are the most current documentation of how processes are conducted [24]. The official documentation for the organization becomes unsynchronized with reality, and limited resources and deadlines encourage the staff to allow the traditional documentation to become stale. Over time, the organization becomes dependent on the software system to answer fundamental questions. The software system effectively becomes the oracle for the organization.

An additional observed side effect of poorly documented legacy systems is that personnel do not easily rotate in and out of organizations charged with maintaining systems with high learning curves [18]. This decreases the competitive flexibility of the organization by having staff either unqualified to perform tasks, or qualified only to perform 1 subset of tasks.

As mentioned before, traditional software engineering often presupposes new development efforts. This is the ideal scenario, but in practice the chance to start systems with a clean sheet of paper happens infrequently. Even if a functional system is to be totally rewritten, using the most current hardware and software available, it is generally replacing a previous, functionally similar system that must be understood, even if it is to be retired. Increasingly, this recognition is evidenced in ways such as self-describing data-files for scientific applications having support for legacy systems explicitly included [8].

The legacy software problem will likely become more prevalent in the research literature, as organizations realize that 1) they have a problem with legacy systems and 2) tools and methods for program understanding are immature. Some examples of reverse engineering support for military legacy systems include [22, 3].

There are 2 significant events accelerating the pace at which legacy systems are being reexamined. The first is the growth of client-server systems. This includes the supporting technologies that are driving this area: the proliferation of cheap, powerful workstation class machines and recent advances in database management systems. This has been an active commercial market for several years. The second and more recent event is the acceptance of World Wide Web (WWW) as a universal interface. At the Fourth World Wide Web conference in Boston, December 1995, there was a significant focus on the application of WWW to legacy systems.

These events and technologies do not directly aid in reverse engineering. Rather, they are responsible for creating a larger demand for reverse engineering tools and techniques. Without client-server proliferation, and to a lesser extent WWW, mainframe codes would have been less likely to receive this sort of wholesale reevaluation.

## Conclusions

Whereas traditional software engineering primarily focuses on "doing it right the first time," reverse engineering addresses the expensive area of maintenance, where one pays the cost of not having done it right the first time, or allowing it to decay over time. Software reverse engineering, or *program comprehension*, is the difficult task of recovering design and other information from a software system. It is difficult to perform because there are intrinsic difficulties in performing the mapping between the language of high level design requirements and the details of low level implementation. Although reverse engineering depends heavily on the human in the loop, there are a variety of automation methodologies available for support. Synchronized Refinement, a hybrid method involving an iterative application of both automation and human intuition and observation, is briefly presented. Finally, the driver for interest in reverse engineering is over 70% of the world's source code is not written in "modern" languages and between 50% and 80% of a systems lifecycle cost is consumed by maintenance. Comprehension of existing systems consumes between 47% and 62% of maintenance resources. Changing hardware, network and information architectures are forcing a revisitation of many legacy systems, particularly mainframe databases. The largest catalyst to date has been the proliferation of client-server computing, which has received renewed energy with the popularity of World Wide Web technologies.

## References

[1] L. A. Belady and M. M. Lehman, "Programming System Dynamics or the Meta-Dynamics of System in Maintenance and Growth," IBM Technical Report RC 3546, September, 1971.

[2] B. W. Boehm, *Software Engineering Economics*, Prentice-Hall, Englewood Cliffs, NH, 1981.

[3] R. Cherinka, C. M. Oversteet, and R. Sparks, "TSME: Maintaining Legacy Systems with Process Driven Software Engineering Technology," Old Dominion University Computer Science Technical Report TR-94-11, January 1994.

[4] E. J. Chikofsky and J. H. Cross, II, "Reverse Engineering and Design Recovery: A Taxonomy," *IEEE Software*, vol. 7, no. 1, pp. 13-17, January 1990.

[5] P. Cousot and R. Cousot, "Abstract Interpretation: A Unified Lattice Model for Static Analysis of Programs by Construction of Appropximation of Fixpoints," Fourth Annual ACM Symposium on Principles of Programming Languages, Los Angeles, CA, January, 1977, pp. 238-252.

[6] J. Ferrante, K. J. Ottenstein, and J. D. Warren, "The Program Dependence Graph and its Use in Optimization," ACM Transactions on Programming Languages and Systems, vol. 9, no. 3, July 1987, pp. 319-349.

[7] R. K. Fjeldstad and W. T. Hamlen, " Application Program Maintenance Study: Report to Our Respondents," Proceedings GUIDE 48, Philadelphia, PA, April 1983.

[8] M. Haines, P. Mehrotra, J. Van Rosendale, "Smartfiles: An OO Approach to Data File Interoperability," ICASE Technical Report 95-56, July 1995.

[9] Maurice H. Halstead, "Elements of Software Science," Elsevier, 1977.

[10] M. Hammer, "Don't Automate -- Obliterate," Harvard Business Review, July/August, 1990, pp. 104-112.

[11] M. Hammer and J. Champy, "Reengineering the Corporation: A Manifesto for Business Revolution," HarperBusiness, New York, 1993.

[12] M. S. Hecht, "Flow Analysis of Computer Programs," North Holland, 1977.

[13] D. Kahaner, C. Moler, and S. Nash, "Numerical Methods and Software," Prentical Hall, Englewood Cliffs, NJ, 1989.

[14] K. Kamper and S. Rugaber, "A Reverse Engineering Methodology for Data Processing Applications," Georgia Tech Technical Report GIT-SERC-90/02, March 1990.

[15] C. McClure, The Three Rs of Software Automation, Prentical Hall, Englewood Cliffs, NJ, 1992.

[16] A. von Meyrhauser and A. M. Vans, "Program Understanding - A Survey," Colorado State University Computer Science Technical Report CS94-120, August 1994.

[17] F. G. Pagan, "Partial Computation and the Construction of Language Processors," Prentice Hall, 1991.

[18] S. Ornburn, and S. Sugaber, "Reverse Engineering: Resolving Conflicts between Expected and Actual Software Designs," Proceedings of the Conference on Software Maintenance, Orlando, Florida, November 1992, pp. 32-40.

[19] D. J. Robson, K. H. Bennet, B. J. Cornelius, and M. Munro, "Approaches to Program Comprehension," *Journal of Systems and Software*, vol. 14, February 1991, pp. 79-84.

[20] S. Rugaber, S. Ornburn, and R. J. LeBlanc, Jr., "Recognizing Design Decisions in Programs," IEEE Software, vol. 7, no. 1, January 1990, pp. 46-54.

[21] S. Rugaber, "Program Comprehension for Reverse Engineering," AAAI Workshop on AI and Automated Program Understanding, San Jose, California, July 1992.

[22] S. Rugaber, and S. Doddapaneni, "The Transition of Application Programs From COBOL to a Fourth Generation Language," Conference on Software Maintenance - 93, Montreal, Canada, September 27-30, 1993.

[23] S. Rugaber, K. Stirewalt, and L. Wills, "The Interleaving Problem in Program Understanding," 2nd Working Conference on Reverse Engineering, Toronto, Ontario, Canada, pp. 166-175, July 14-16 1995.

[24] S. Rugaber, "Program Comprehension," *Encyclopedia of Computer Science and Technology*, Draft -- to appear, April, 1995.

[25] B. Shneiderman, "Software Psychology: Human Factors in Computer and Information Systems," Little Brown, 1980.

[26] Linda M. Wills, "Using Attributed Flow Graph Parsing to Recognize Programs," Workshop on Graph Grammars and Their Application to Computer Science, Williamsburg, Virginia, November 1994.

[27] Edward Yourdon, "Structured Walkthroughs," Yourdon Press, 1989.